# *Ab Initio* Nucleic Acid Folding Simulations Using a Physics-Based Atomistic Free Energy Model

Chi H. Mak

Departments of Chemistry and Quantitative and Computational Biology, and Center of Applied Mathematical Sciences, University of Southern California, Los Angeles, California 90089, USA

To whom correspondence may be addressed:

Chi H. Mak, Department of Chemistry, University of Southern California, Los Angeles, CA 90089
Tel: 213-740-4101, Fax: 213-740-3972, E-mail: cmak@usc.edu

## Abstract

Performing full-resolution atomistic simulations of nucleic acid folding has remained a challenge for biomolecular modeling. Understanding how nucleic acids fold and how they transition between different folded structures as they unfold and refold has important implications for biology. This paper reports a theoretical model and computer simulation of the *ab initio* folding of DNA inverted repeat sequences. The formulation is based on an all-atom conformational model of the sugar-phosphate backbone via chain closure, and it incorporates three major molecular-level driving forces – base stacking, counterion-induced backbone self-interactions and base pairing – via separate analytical theories designed to capture and reproduce the effects of the solvent without requiring explicit water and ions in the simulation. To accelerate computational throughput, a mixed numerical/analytical algorithm for the calculation of the backbone conformational volume is incorporated into the Monte Carlo simulation, and special stochastic sampling techniques were employed to achieve the computational efficiency needed to fold nucleic acids from scratch. This paper describes implementation details, benchmark results and the advantages and technical challenges with this approach.

# I. Introduction

Biomolecular simulations of nucleic acids, DNA and RNA, are often carried out using one of two methods[1–5]. In molecular dynamics (MD), atoms on the nucleic acid molecule are represented by classical particles held together by bond, bond angle and torsion potentials, and forces from these potentials and non-bonded interactions are used to advance Newton's equation. While MD can be performed with implicit solvent, explicit water molecules are necessary to correctly model most biomolecules. In coarse grain (CG) methods[2–6], a nucleic acid is often represented by several united atoms, typically for the sugar, the base and the phosphate, to accelerate the propagation of Newton's equation. In general, the parameters of the CG force field must be tuned to suit different applications. Their transferability is thus limited compared to all-atom MD. Depending on what questions are to be addressed, the missing atomistic details in CG models can be a deficit.

Both MD and CG methods can be used for simulating the folded structure of nucleic acids, but if simulating the details and thermodynamics of the molecular processes behind nucleic acid folding, misfolding, unfolding or refolding is the objective, both MD and CG have intrinsic limitations. MD has a speed problem, because a fully atomistic nucleic acid molecule moves through conformational space slowly. But even if the nucleic acid does move quickly, any large-scale conformational motion of the nucleic acid itself will be hindered by explicit solvent molecules. CG also suffers from a lack of first-principle accuracy. While the solvent comprising of waters[7,8] and counterions[9–11] ($Na^+$, $K^+$ or $Mg^{2+}$) is essential for modeling the backbone and the base-pairing and base-stacking interactions that drive nucleic acid structures, the predictability of CG models without explicit solvent is nonuniform across nucleic acids with different nucleotide sequences.

This paper considers an alternative to MD and CG for the simulation of *ab initio* nucleic acid folding. The method described here is based on Monte Carlo (MC) and has the atomistic accuracy of MD, but the MC moves that propel the nucleic acid molecule through its conformational space are designed to have intrinsically higher efficiency than MD. Accurate free energy models are combined with this method to capture the effects of waters and counterions on the base-base and backbone-backbone interactions, providing a solvent-free simulation, and at the same time, retaining first-principle accuracy.

Various theoretical elements at the core of this method have been presented previously. This paper describes how they are integrated into a rigorous equilibrium simulation in order to be able to fold nucleic acids, starting from an initially unfolded configuration *ab initio* (i.e. without hardcoding any specific base pairing information into the sequence). This method provides a rigorous equilibrium simulation strategy, enabling the statistical mechanics of nucleic acid folding to be studied computationally from first principle. This method has also been optimized numerically to be able to fold nucleic acid sequences ~30 nucleotide (nt) with reasonable computational resources (~2 weeks CPU time). This paper presents a benchmark of this method for the folding of an inverted repeat (IR) DNA oligonucleotide sequence. IR sequences on the genome are known to be able to form alternative secondary structures such as cruciforms[12–16]. IRs serve many biological functions such as recognition sequences at replication origins[16,17] or for recombination[18–21]. IRs have also been implicated as one of the potential factors leading to genome instability[22–25]. The results will show that with an efficient sampling method and with the physics of the solvent captured by viable theoretical free energy models, *ab initio* folding of nucleic acids can be achieved within an atomistically accurate simulation.

The basis of the new MC method is the chain closure algorithm[26–32]. Chain closure (CC) is mathematically quite different from MD. In CC, the backbone of a linear polymer is viewed as atoms linked by (approximately) rigid bond lengths and bond angles, and the flexible dihedral angles along the backbone then give rise to conformational fluctuations of the polymer. The resulting conformational volume $\Omega$ of the backbone corresponds to the chain entropy $S$, via the Boltzmann formula $S = k_B \ln \Omega$. Because the molecular interactions that drive nucleic acid folding, such as base stacking and base pairing, are established in coordinate space, a direct MC sampling in dihedral angle conformation space by sampling the torsions along the sugar-phosphate backbone of a nucleic acid is numerically inefficient. A better strategy is to look for an alternative set or sets of dihedral angles to reclose the backbone in conformation space when base stacking and base pairing interactions are sampled in real space, and CC MC algorithms are implementations of this mathematics. The polymer backbone conformation problem is equivalent to a robotic assembly with six consecutive revolute joints, which has been studied in robotics as the so-called 6R problem[30]. Using inverse kinematics, the closure problem has been solved for the sugar-phosphate backbone[33–35], but the solution is numerically costly. In practical MC implementations of CC, different levels of approximation to the closure solution to various degrees of accuracy are usually employed[31,32]. But while CCMC simulations can more efficiently sample conformational fluctuations along the nucleic acid backbone, they typically do not include explicit water because the vast majority of backbone conformational rearrangements are rejected due to excluded volume clashes with explicit water, negating the computational advantage coming from CC. But because the aqueous environment is an essential effector of many of the key interactions driving nucleic acid structure formation, CCMC without explicit solvent is nonpredictive, and its utility has been largely confined to simulating double-stranded DNA[31,36,37] where large-scale conformational changes are muted. To date, CC has not yet been shown to be effective for studying folding or any secondary structural transitions in nucleic acids. This paper presents a different and viable implementation of CC for folding nucleic acids.

## II. Theoretical Model

This section describes key components of the theoretical model behind the simulation method. Elements of some of these have been reported previously and details will not be reproduced here. We only focus on those aspects of the theories most relevant to the simulation, as well as any that are new. The four fundamental physical forces that have been identified by the CCMC simulations as essential for folding are: (1) backbone conformational entropy, (2) counterion-induced intrachain attractions, (3) base stacking interactions and (4) base complementarity interactions, and these constitute the four key terms in the model free energy cost function described in this section.

### A. Conformational entropy of the nucleic acid backbone

As discussed in the last section, CC facilitates an efficient sampling of the conformational fluctuations along the nucleic acid backbone. The dominant nucleotide-specific interactions that define the folded structure of a nucleic acid are primarily due to the bases. Stacking and base pairing forces between the bases are two key factors the algorithm must contend with, so an efficient sampling algorithm should focus on sampling the positions of the bases, while the conformational fluctuations of the backbone atoms must then comply with these base positions. For example, when two sequence-neighbor bases are stacked against each other, instead of trying to update the backbone dihedral angles along the chain between these two bases to a single new configuration, it will be much more productive if one can enumerate *all* the conformations that are consistent with the positions of these two stacked bases (or compute them analytically), since the total conformational volume of the backbone atoms between the two bases defines an entropic cost to the backbone. If the total conformational volume of all dihedral angle configurations of the backbone can indeed be computed analytically instead of by sampling one backbone conformation at a time, the efficiency of the sampling will be greatly enhanced. The key strategy in CCMC is to update the positions of the bases, but analytically compute the change in the backbone entropy by calculating the total conformational volume of all dihedral angle configurations along the entire chain permitted by the new positions of the bases versus that permitted by the old positions. A single base can be moved, or many bases can be moved together to form a trial move, but the backbone entropy can be recalculated the same way.

The CC procedure for the sugar-phosphate backbone is illustrated in Fig. 1. Details have been described in previous papers[33–35,38]. Fig. 1(a) shows a segment of the chain with four nucleotides. The $j$-th and $j+1$st nucleotides have been labeled. Only heavy atoms are explicitly shown, and two of the phosphate oxygens have been omitted for clarity. The glycosidic N atoms ($N_1$ for pyrimidines, $N_9$ for purines) are shown in blue, the anomeric $C_{1'}$ atoms in green and oxygens in red. One phosphate group connects the two sequence-neighbor nucleotides $j$ and $j+1$. Fig. 1(b) shows an expanded view of the $j+1$st nucleotide, illustrating the base and the deoxyribose sugar unit. Various formulations of closure for nucleic acids are found in the literature[31,32,38]. In our formulation, the three atoms labeled "$CN_bC$" serve key functions – they are the "base carbon" ($C_2$ for pyrimidines, $C_4$ for purines), the glycosidic $N_b$ atom and $C_{1'}$. Because the glycosidic bond $C_{1'}$-$N_b$ plays a special role, it is shown in white in Fig. 1.

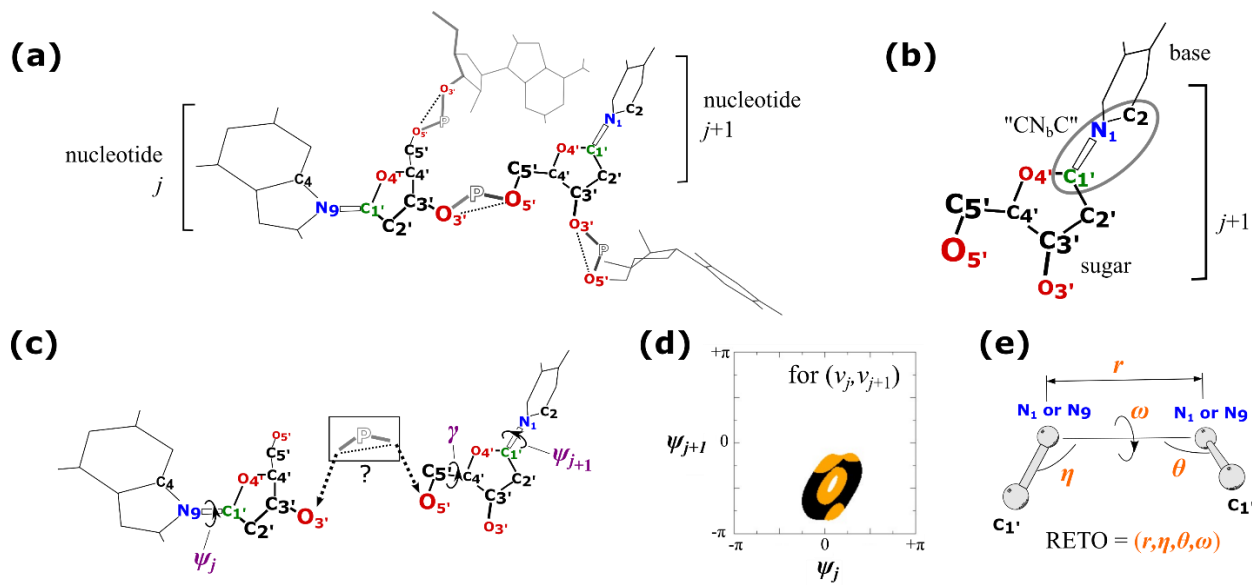

**FIGURE 1.**

Chain closure for the nucleic acid backbone. (a) Schematic of a portion of the sugar-phosphate chain showing four nucleotide units, and two sequence neighbors labeled $j$ and $j+1$. The $C_4$, $N_9$ and $C_{1'}$ on nucleotide $j$ are defined as the $(CN_bC)_j$ atoms, and the $C_2$, $N_1$ and $C_{1'}$ atoms on nucleotide $j+1$ are $(CN_bC)_{j+1}$. Specifying the position and orientation of the $CN_bC$ atoms for every nucleotide completely determines the coordinates of all the base atoms. (b) Given coordinates of the $CN_bC$ atoms of nucleotide $j+1$ and a particular puckering state of the ribose ring, coordinates of all the ribose atoms are completely determined by dihedral angle $C_2$-$N_1$-$C_{1'}$-$C_{2'}$, which specifies the rotation of the sugar around the $N_1$-$C_{1'}$ bond on the $CN_bC$ group. The same is true for nucleotide $j$ (not shown). (c) For every pair of these ribose rotation angles ($\psi_j$, $\psi_{j+1}$), the chain is reclosed by inserting the phosphate group to connect the two sugars. The dihedral angle $O_{4'}$-$C_{4'}$-$C_{5'}$-$O_{5'}$ $\gamma$ is determined as the solution to the closure problem. (d) Depending on whether the closure solution exists between nucleotides $j$ and $j+1$, the viable solutions are represented as the shared (black+orange) domains on the ($\psi_j$, $\psi_{j+1}$) plane. Collisions between the two ribose units may exclude additional regions, e.g. those shaded in orange. The remaining regions in black are the final viable closure domains. (e) The four independent internal coordinates ($r, \eta, \theta, \omega$) specifying the relative position and orientation between the $N_b$-$C_{1'}$ bond on one $CN_bC$ group and the next.

As discussed above, the structure of a folded chain is primarily defined by where base pairs are made and how the bases are stacked. The most efficient sampling scheme would be one where the positions of the bases are moved or sampled by MC, while all the conformational states of the sugar-phosphate backbone before and after each move can be enumerated on-the-fly to evaluate the entropy change. This enumeration is illustrated by Fig. 1(c). The two nucleotides $j$ and $j + 1$ are shown in Fig. 1(c) with the phosphate unit between them disconnected. With the bases at their old positions before a MC move, the conformational volume of the backbone between them needs to be calculated, and this is repeated for the new base positions. Because the bases are planar (except for the methyl group on thymines), knowing the positions of the bases is equivalent to specifying the coordinates of the $CN_bC$ atoms on both $j$ and $j + 1$. With the phosphate group disconnected, the sugar unit on each nucleotide is free to rotate around each $N_b$-$C_{1'}$ bond shown in white in Fig. 1(c). $\psi_j$ and $\psi_{j+1}$ are used to denote these two rotation angles. The next subsection will give prescription for how the conformational states of each deoxyribose ring can be enumerated too, but for the time being it is easy to see that the different conformational solutions of the ribose correspond to different puckering states of the sugar, and the conformation of the ring can be precomputed for each. Each of these puckering geometries is controlled by a variable $\nu^*$, and the corresponding ring atoms with its conformation specified by $\nu^*$ can be ligated to the $CN_bC$ unit at the $C_{1'}$ atom. This is done separately for nucleotides $j$ and $j + 1$, and it results in the two disconnected nucleotides shown in Fig. 1(c).

The two rotation angles $\psi_j$ and $\psi_{j+1}$ are free, but only specific combinations will properly reclose the chain. Their values will be determined as solutions to the closure problem. Notice that the $5'$ and $3'$ directions are nonequivalent, and nucleotide $j$ up to the atom $O_{3'}$ are fixed by choosing a certain ring conformation $\nu_j^*$ and a specific $\psi_j$, whereas nucleotide $j + 1$ has an extra $O_{5'}$ atom after the phosphate, so there is one extra degree of freedom on the right side in addition to the ring conformation $\nu_{j+1}^*$ and rotation angle $\psi_{j+1}$. This extra degree of freedom corresponds to a rotation angle around the $C_{5'}$-$C_{4'}$ bond on nucleotide $j + 1$, which is shown as $\gamma$ in Fig. 1(c). Specifying the $\gamma$ angle fixes the coordinates of $O_{5'}$. The closure solution is therefore given by the set(s) of $(\psi_j, \psi_{j+1}, \nu_j^*, \nu_{j+1}^*, \gamma)$ that would allow the backbone to be reclosed properly when the phosphate is reinserted to connect the chain back up.

Before going to the final solution, which is illustrated in Fig. 1(d), notice that there are five degrees of freedom $(\psi_j, \psi_{j+1}, \nu_j^*, \nu_{j+1}^*, \gamma)$ but the problem has four input variables. These input variables are the internal coordinates that describe the position of the $C_{1'}$-$N_b$ bond on nucleotide $j + 1$ relative to the $C_{1'}$-$N_b$ bond on nucleotide $j$. These internal coordinates are the virtual bond distance between the two $N_b$ atoms, the two virtual bond angles with the $C_{1'}$ atoms, and one dihedral, labeled $(r, \eta, \theta, \omega)$ in Fig. 1(e). We refer to these internal coordinates collectively as "RETO", and the set between nucleotides $j$ and $j + 1$ is denoted $RETO_j$.

Closure is completed by inserting the phosphate group back in to bridge the two nucleotide units. The O-P-O unit has a definite O…O distance because the two P-O bonds and the O-P-O angle are fixed, so the condition introduced by recoupling the phosphate is equivalent to a $O_{3'}…O_{5'}$ distance constraint. The solution is specified by domain(s) on the $(\psi_j, \psi_{j+1})$ plane where the chain can be properly reconnected between nucleotides $j$ and $j+1$ once a ring conformation has been chosen for each of $\nu_j^*$ and $\nu_{j+1}^*$, and the degree of freedom $\gamma$ is solved to satisfy the $O_{3'}…O_{5'}$ distance constraint. The solution is illustrated pictorially in Fig. 1(d), where the shaded areas on the $(\psi_j, \psi_{j+1})$ plane represent domains allowed by closure, while the rest does not. Within the shaded regions on the $(\psi_j, \psi_{j+1})$ plane, not all are physically accessible because some of these solutions may incur collisions between atom(s) on nucleotide $j$ with atom(s) on $j+1$. Fig. 1(d) illustrates these collision-excluded regions in orange. After these are removed, the domains colored in black represent the remaining viable solutions. For each pair of ring conformations $(\nu_j^*, \nu_{j+1}^*)$, there is a different closure solution; therefore, one can think of these closure solutions as $C(\psi_j, \psi_{j+1}; \nu_j^*, \nu_{j+1}^* | \text{RETO}_j)$ where $C = 1$ inside closure- and collision-allowed domains, and 0 otherwise.

To compute the total conformational volume for the entire chain, $C$ is integrated over all nucleotide units:

$$\Omega = \int d\psi_1 \, d\nu_1^* \int d\psi_2 \, d\nu_2^* \cdots \int d\psi_N \, d\nu_N^* \prod_1^{N-1} C(\psi_j, \psi_{j+1}; \nu_j^*, \nu_{j+1}^* | \text{RETO}_j) \qquad (1)$$

yielding the conformational entropy of the chain as $S = k_B \ln \Omega$, where $k_B$ is Boltzmann's constant. As the positions of the bases are sampled during MC, $\Omega$ is calculated before the move and for the trial move, to accept or reject according to detailed balance. Since Eq. (1) is an integral, the entire integration must be redone every time any base is moved, so this calculation can be costly. In practice, segments that are far from each other along the chain become uncorrelated quickly. In the simulations, the change in $\Omega$ is calculated by carrying the integration in Eq. (1) up to some cutoff from the base that is being moved, which is typically about 6 nucleotides away, beyond which their correlations are close to zero.

While conceptually simple, numerical implementation of the closure solution is somewhat involved. The technical details will be described in a later section. Here, several features of this CC solution should be pointed out. First, the method as described above is not a true implementation of the original 6R solution[34,39]. In the 6R problem, six consecutive dihedral angles along the chain comprise the degrees of freedom, and the proper solution preserves all bond lengths as well as the bond angles. Exact solutions of the 6R problem consist of a set of discrete points. The theoretical maximum number of solutions between each nucleotide pair is 64, but the actual number of solutions is usually far fewer. Since the puckering of the ribose is also coupled to the rest of the dihedral angles around the ring, one of the six dihedral angles along the sugar-phosphate backbone in each nucleotide unit is not strictly free, and this further reduces the number of viable solutions. In practice, the rigid bond-length/bond-angle assumption behind the 6R problem is never rigorously obeyed by polymers. Instead of implementing the 6R problem, the closure solution described above solves an approximation of it, known as RSSR (a system with two revolute and two spherical joints)[40]. In the RSSR problem, not all bond angles are preserved. In Fig. 1(c), one can easily see that the non-preserved angle is either $C_{3'}$-$O_{3'}$-P or P-$O_{5'}$- $C_{5'}$. In fact, high-resolution X-ray structures show that many nucleic acids display very diverse values for their $C_{3'}$-$O_{3'}$-P and P-$O_{5'}$- $C_{5'}$ angles. The RSSR solution, while being an approximation to the 6R problem, is not only numerically better behaved, it also relaxes the rigid bond-length/bond-angle assumptions in 6R closure. To improve the accuracy of the model, additional empirical potentials were added to control the persistence along the backbone as well as the right-handed twist of the chain. With the persistence and twist measures defined in Fig. 2, a persistence bias of $(3.6 \text{ kcal/mol}) \sum_j p_j$ and a twist bias of $(-0.70 \text{ kcal/mol}) \sum_j \cos(t_j)$ were added to the backbone free energy.

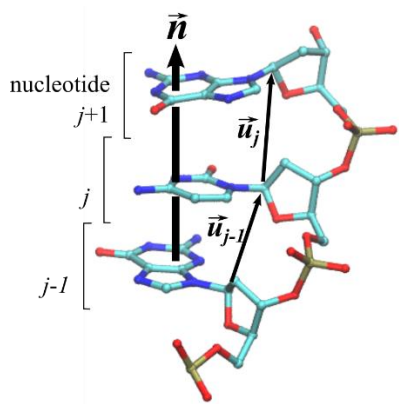

**FIGURE 2.**

For three consecutive nucleotides, $j-1, j, j+1$, along the sugar-phosphate backbone, $\vec{u}_{j-1}$ and $\vec{u}_j$ are defined to be the two unit vectors from $C_{1'}$ of $j-1$ to $j$ and from $j$ to $j+1$. The persistence of the chain at $j$ is defined as $p_j \equiv \vec{u}_{j-1} \cdot \vec{u}_j$. The consensus unit normal vector to the three bases is $\vec{n}$, and the plane perpendicular to $\vec{n}$ is defined as $N$. The twist $t_j$ at $j$ is defined as the angle made by the projections of $\vec{u}_{j-1}$ and $\vec{u}_j$ onto plane $N$.

## B. Deoxyribose ring conformations

The mainchain closure solution in the last section assumes that all ribose ring conformations can be precomputed, and each conformation is characterized by a variable $v^*$. This section describes how these ribose ring solutions were obtained. Details can be found in the original papers[26,27,33].

Around the five-atom ring of each ribose, there are five internal dihedral angles, customarily denoted $v_0$ to $v_4$. Once these dihedral angles are specified, together with the known stereogeometry around each of the carbon atoms on the D-deoxyribose, absolute coordinates of all the sugar atoms except $O_{5'}$ can be determined. As described in the last section, further specifying $\gamma$ also defines the coordinates of $O_{5'}$.

Since the ribose has a ring structure, the angles $\nu_0$ to $\nu_4$ can be determined by closure like in a linear chain. But since there are only five ring dihedral angles, they do not satisfy the condition for a 6R problem. At least one of the bond angles cannot be strictly conserved. So instead of solving the ribose ring closure as a 6R problem, it is solved as a RSSR problem similar to the closure of the backbone. In the RSSR solution, only one of the five ring dihedral angles is free, and this is the variable referred to in the last subsection as $\nu^*$, which corresponds to the $\nu_1$ angle in previous work[33]. Additionally, there are up to two possible ribose ring closure solutions for each $\nu^*$, corresponding to two different puckering states[27]. All of these ribose ring closure solutions can be solved ahead of the simulation. The results, including the coordinates of all the ribose atoms as a function of $\nu^*$, can be precomputed, stored and reused throughout the simulation.

## C. Free energy for counterion-induced self-interactions along the chain

The sugar-phosphate backbone of a nucleic acid is a polyelectrolyte. Under physiological conditions, one unit of negative charge is associated with every phosphate. But the sugar-phosphate backbone exhibits attractive intrachain self-interactions when dissolved in an aqueous solution in the presence of cations ($Na^+$, $K^+$ or $Mg^{2+}$)[9,41]. Experimentally, these counterion-induced interactions are observed to stabilize folded nucleic acid chains compared to open unfolded ones in solutions with monovalent counterions such as $Na^+$ or $K^+$ in the concentration range of 0.1 to 1 M, or those with divalent counterions such as $Mg^{2+}$ in the range of a few mM[10]. Previous computational studies[11,42] showed that without these counterion-induced self-interactions, nucleic acids do not achieve stable folds even when the charges on the phosphates are removed. A proper counterion free energy must therefore be added to the theoretical model for it to successfully model folding. But including explicit counterions in the simulation creates a different sampling problem, because explicit counterions equilibrate very slowly. To properly account for counterion effects, these self-interactions along the chain are captured by an analytical free energy model in the CCMC simulations.

Details of this counterion free energy model have been given in previous papers[42–44]. Briefly, the theory assumes divalent counterions such as $Mg^{2+}$ are bound to the phosphates in a tight-binding model, and the rest of the counterions around the polymer form a diffuse charge cloud, the effects of which are treated by a classical density functional theory. For every conformation of the sugar-phosphate backbone, the locations of the phosphate groups are defined. Let these be $\{\boldsymbol{R}_1, \boldsymbol{R}_2, \cdots, \boldsymbol{R}_N\}$. In the tight-binding model, a phosphate at $\boldsymbol{R}_i$ either has a $Mg^{2+}$ ion bound to it, in which case the total charge on that phosphate is $\sigma_i = +1$, or has no $Mg^{2+}$, in which case the charge of the bare phosphate is $\sigma_i = -1$. Mean-field solution of the density functional theory for the diffuse ions produces a Debye-like screening on the electrostatic interactions between the phosphate tight-binding sites. In this theory, the free energy induced by the counterions on the nucleic acid, $F_{CI}$ is simply given by:

$$e^{-\beta F_{CI}} = \mathrm{Tr}_{\{\sigma_i\}} \exp\left(-\beta \sum_{i<j} \frac{\sigma_i \sigma_j}{|R_i - R_j|} \frac{e^{-|R_i - R_j|/\Lambda_{CI}}}{\epsilon_{CI}}\right) \tag{2}$$

where $\beta = 1/k_B T$ and Tr is a trace over all possible configurations of $\{\sigma_i = \pm 1\}$, with the constraint that their sum adds up to zero, which is true for typical counterion concentrations where the charge on the nucleic acid chain is overall neutral. In essence, Eq. (2) is just the partition function for a 1-d Ising model with interactions that are functions of the current positions of the phosphates in the simulation. The Debye screening length $\Lambda_{CI}$ and the counterion dielectric $\epsilon_{CI}$ in Eq. (2) are adjustable parameters in this theory. Previous calculations suggest that $\epsilon_{CI}$ in the range of ~ 20 to 12 is consistent with the effects of $Mg^{2+}$ in the concentration range ~ 1 to 5 mM[44]. During the CCMC simulation when the positions of the bases are sampled and the backbone conformational entropy calculated using closure, the trace in Eq. (2) is carried out simultaneously by a numerical search routine that keeps track of the lowest-energy configurations of the counterions $\{\sigma_i\}$ (~30 configurations), using these states on the lowest end of the spectrum to approximate the free energy $F_{CI}$ produced by the counterions on the backbone according to Eq. (2).

## D. Free energy for solvent-induced base stacking interactions

Stacking is another major determinant of nucleic acid structure. A recent fully atomistic explicit water simulation from our group demonstrated that the molecular effect that is commonly known as "base stacking" appears to be largely due to the aqueous solvent[7]. Purine and pyrimidine bases are planar, and when one base is stacked against another, a number of water molecules in the solvent must be excluded from the interface between the two bases. Depending on their distance and how they stack, these two bases exclude different number of water molecules, and it is the entropic loss felt by the aqueous solvent due to the exclusion of these water molecules that generates a propensity between the two bases to stack against (i.e. parallel to) each other. A cartoon of these exclusion effects is shown on the left side of Fig. 3, where two stacked bases are shown to exclude water molecules from the volume between them when their spacing is less than the diameter of one water. This imposes an entropic penalty on the solvent. The solvent tries to push the two bases back together in an attempt to minimize its entropy loss, and this is largely the origin of the stacking free energy. The cartoon in the middle of Fig. 3 shows that this entropic loss is partially recovered when the distance between the two bases can accommodate a single layer of water molecules. Validating these ideas, Fig. 3 shows simulation results[7] of the solvent's contribution to the stacking interaction between the two bases as a function of their separation, revealing an oscillatory dependence which reflects the underlying solvent structure on an atomistic length scale[45].

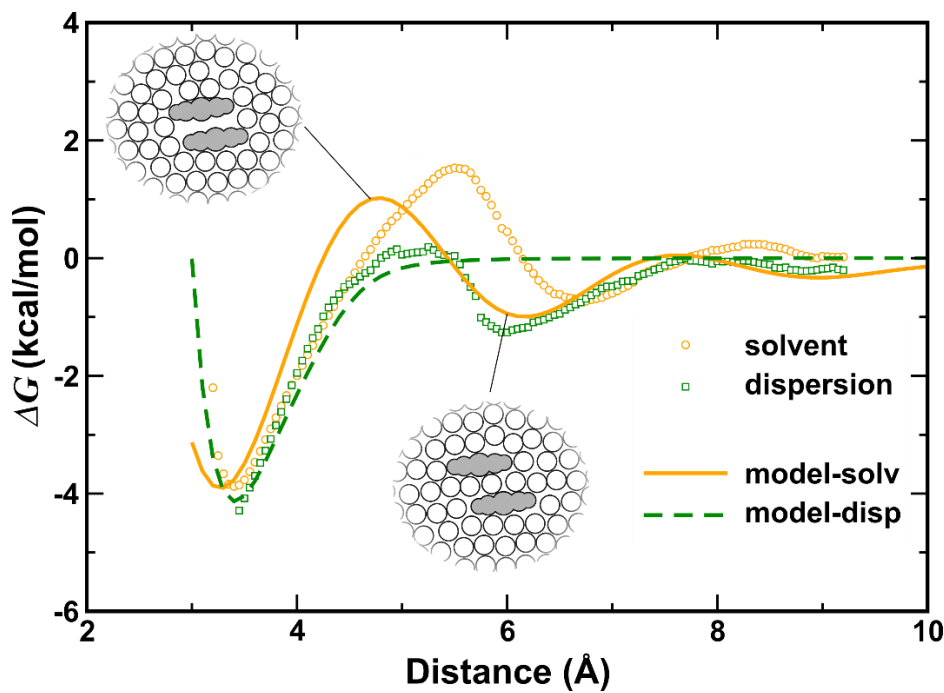

**FIGURE 3.**

Results from all-atom umbrella free energy simulations[7] of two parallelly stacked guanines in water showing the entropic part of the free energy between them (orange circles) and the solvent-renormalized dispersion interaction (green squares) as a function of separation. The corresponding free energies from the theoretical model described in [46] are shown as the orange solid line (solvent entropic term) and the green dashed line (solvent-renormalized dispersion).

Informed by these simulation data, an analytical theory for the entropic loss incurred by the solvent was formulated[46] based on the work of Pratt et al.[47–49] and Chandler[50,51]. Fig. 3 shows the solvent's contribution to the stacking free energy from the simulations between two guanines as the orange circles and the dispersion interaction between the two bases as the green squares, as a function of their separation when they are constrained to remain parallel to each other, and the yellow solid line in displays the solvent-induced free energy from this analytical theory, with the green dashed line showing the solvent-renormalized dispersion interaction between the bases. The oscillatory behaviors of these functions reflect the solvent's central role[45] in the stacking interactions. This stacking free energy model was incorporated into the CCMC simulations. While the analytical theory is able to reproduce the essential physics and the approximate strengths of base stacking interactions for two stacked bases in an aqueous environment, Fig. 3 shows that the precise distance-dependence of the theory (solid yellow line) does not perfectly match results from the all-atom simulations (yellow circles), but the stacking distance of ~ 3Å as well as the characteristic oscillatory distance-dependence of the stacking free energy are correctly reproduced.

## E. *Ad hoc* model for solvent-renormalized base complementarity interactions

Base pairing interaction is the final essential physical force. Watson-Crick (WC) pairing is the basis for complementarity, but while WC interactions are thought to be predicated on hydrogen bonding in G|C and A|T pairs, the strength of these base pairing interactions are heavily renormalized by the aqueous solvent around them. A recent umbrella free energy calculation in explicit water[8] shows that a G|C pair has an interaction strength that is surprisingly similar to an A|T pair even though there are thought to be three hydrogen bonds between G|C and only two in A|T. The numerical results suggest that the strength of WC base pairs is of the order of ~ −6 to −8 kcal/mol (negative free energy is stable), but the difference between G|C and A|T is only ~ 1 or 2 kcal/mol. Fig. 4 shows base pair free energies as a function of hydrogen bonding donor-acceptor distance as the two bases are separated from each other while constrained to remain coplanar. The A|T base pair free energy is shown as the closed orange circles, and G|C pair as closed green squares. In the same simulation, the direct hydrogen bonding interactions (before they were renormalized by water) were also computed, and they are shown in Fig. 4 as the open orange circles for an A|T pair and open green squares for a G|C pair. Clearly, the direct base pairing interactions have been heavily renormalized in water.

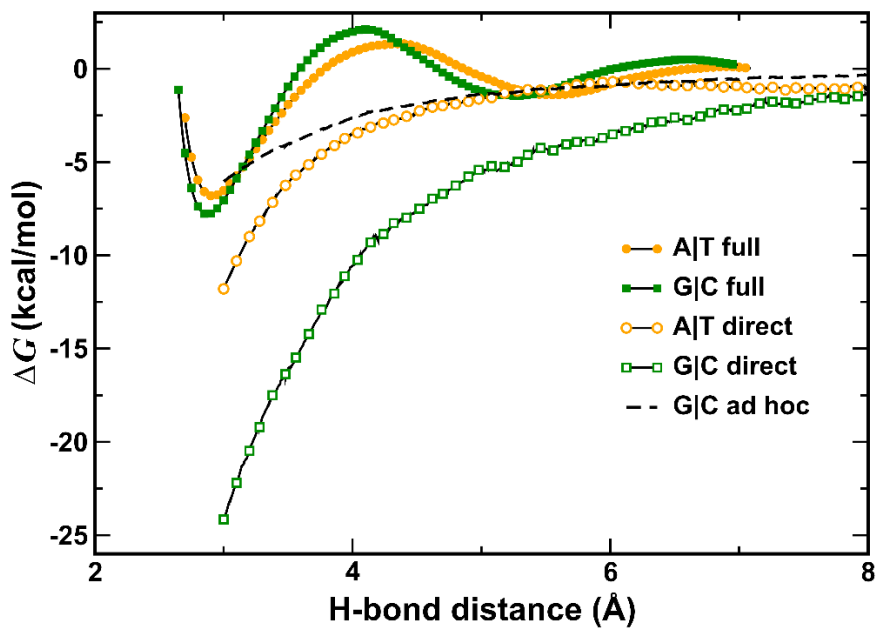

**FIGURE 4.**

Results from all-atom umbrella free energy simulations[8] of canonical base pairs in water. The overall pairing free energy for A|T in water is shown as closed orange circles, and for G|C as closed green squares. In solvent, the pairing free for G|C is only ~ −1 kcal/mol deeper than A|T. Direct hydrogen bonding interaction energy, which is the full free energy minus solvent's contribution, are displayed as

open orange circles for A|T and as open green squares for G|C. The *ad hoc* model used in the simulation for G|C, whose attractive part is the direct interaction divided by $\epsilon_{GC}$, is shown as the dashed line.

Attempts are ongoing to formulate a free energy model for the solvent-renormalized base pairing interactions. But in lieu of a theory, an *ad hoc* model had been adopted in the CCMC simulations. As a proxy for the fully solvent-renormalized base pair free energies, the direct interactions were attenuated and used as a model for the pairing interaction. Assigning partial charges to the base atoms according to Amber ff94[52], charge-charge interactions in a G|C or A|T pair were used to approximate the solvent-renormalized base complementarity interactions. Based on the data in Fig. 4, the proper dielectric constant to use for a G|C pair was found to be $\epsilon_{GC} \sim 4$, whereas for a A|T pair, $\epsilon_{AT} \sim 2$. (These empirical dielectrics, $\epsilon_{GC}$ and $\epsilon_{AT}$, were used in the *ad hoc* pairing interactions to attenuate the direct interactions, while the counterion-induced free energy term employed its own dielectric $\epsilon_{CI}$. Besides these, no other free energy term in the model involved charges.) In addition to empirically attenuating the strength of the electrostatic energy in each base pair in order to use it as a proxy for the pairing free energy, it was also necessary to adjust these interactions according to the coplanarity of each base pair $(j, k)$ by an additional empirical scaling factor $|\vec{n}_j \cdot \vec{n}_k|$, where $\vec{n}_j$ denotes the unit normal vector to the plane of base $j$. This factor de-emphasized pairing when two bases approach each other with a T geometry but favors geometries that are near coplanar. While this *ad hoc* base pairing model does not precisely reproduce the details of how the water molecules renormalize the direct interaction between two paired bases, the folding simulation results demonstrate that this *ad hoc* pairing model contains sufficient physics to properly describe base complementarity in G|C pairs in the context of folding. However, the same *ad hoc* model failed for A|T pairs, and the reasons will be discussed with the folding results below.

# III. Monte Carlo Simulations and Other Algorithmic Details

This section provides computational details of how the theoretical models in Sect. II have been implemented in CCMC simulations.

## A. Excluded volume between atoms and other atomic parameters

In addition to the free energy terms described in Sect. II, every pair of atoms in the model also has an excluded volume interaction between them to represent their steric repulsion. For this, Lennard-Jones (LJ) parameters were adopted from Amber ff94[52], employing the Weeks-Chandler-Andersen (WCA) potential[53] instead of LJ to capture the steric forces. These excluded volume interactions were enforced between all atoms in the model, on the backbone as well as the bases.

For the counterion free energy model described in Sect. II C, a $-1$ charge was assigned to each phosphate group. As described in Sect. II C, when a $Mg^{2+}$ ion is bound to a phosphate group, its charge turns into $+1$, and the tight-binding free energy Eq. (2) was used to compute the counterion-induced self-interactions between the chain and itself. Other than the phosphate group, all other atoms on the backbone were assigned no charge. These phosphate charges were only used in the calculation of the counterion-induced free energy. They were not used for any other term in the free energy cost function.

For the calculation of the solvent-induced stacking interactions described in Sect. II D, each base unit was assigned a space-filled volume according to the prescription in [46]. This space-filled volume was calculated using the WCA parameters of the atoms on each of the bases. The solvent-entropy-driven stacking interactions were calculated between every pair of bases in the model, using the recipe in[46]. In addition to the solvent-entropy-driven stacking terms, a solvent-renormalized dispersion interaction between sequence-neighbor bases was also included based on the attractive branch of the LJ interactions between atoms on different bases. Details are given in [46].

Finally, as described in Sect. II E, the pairing interaction between every pair of bases was modeled by a Coulombic energy between partial charges centered on each base atom, renormalized by an empirical dielectric constant. The atomic partial charges on the bases were taken from Amber ff94. Partial charges were assigned only to atoms on the bases, and backbone atoms had no partial charges so there were no electrostatic interactions between bases and backbone. Since the backbone atoms are assumed neutral, all excess charges from the atoms on each base were offset by an opposite charge assigned to its $N_b$ atom. Coulombic interactions were computed for every pair of bases according to the prescription in Sect. II E.

## B. Sampling the equilibrium ensemble

A single DNA in the canonical ensemble was simulated at a fixed temperature of 310 K. The effects of the aqueous solvent on the DNA were described by the models in Sect. II D and E, and counterions corresponding to ~ 2mM $Mg^{2+}$ were described by the model in Sect. II C with $\Lambda_{CI} = 140\text{Å}$. The $(CN_bC)_j$ atoms of each nucleotide $j$ were used in the simulation to keep track of the position and orientation of the base on each of the nucleotides. Since there are no internal degrees of freedom on the bases, specifying the positions and orientations of all $(CN_bC)$ groups also completely defines the coordinates of the atoms on all the bases. The conformational entropy of the sugar-phosphate backbone between these $CN_bC$ atom groups was calculated using the prescription in Sect. II A. The chain's self-interaction free energy due to the counterions was calculated according to the recipe in Sect. II C. The stacking and pairing interactions between every pair of bases were computed according to the methods in Sect. II D and E.

Several types of MC moves were employed. The energy before and after each trial move were computed on the fly, and the trial move was accepted or rejected based on the Metropolis algorithm[54]. Since each $(CN_bC)_j$ group is internally rigid (bond angles and bond lengths are fixed), the most rudimentary MC move consisted of applying random translation and rotation to each $(CN_bC)_j$ group on the DNA in a random order. One round of this basic translation+rotation move carried out on every $(CN_bC)_j$ group constitutes one MC step (MCS).

Single $(CN_bC)$ translation+rotation moves became progressively inefficient as more base pairs and base stacks develop in the structure, since stacking and pairing stabilize increasingly larger-scale structures. To achieve higher ergodicity as the chain folds, additional MC moves aimed at effecting larger-scale conformational rearrangements on the positions of the $(CN_bC)$ groups were employed. Single translation+rotation moves on a randomly selected set of $(CN_bC)$ atom groups either on the 5′ or the 3′ end of the chain was added during each pass. These chain-end moves helped accelerate conformational rearrangements of the sequence, predominately at the initial stages of folding. In addition to moving a set of $(CN_bC)$ atoms groups at either ends of the chain, any randomly selected set of $(CNbC)$ units belonging to a subsequence on the interior of the chain can also be moved in the same way, but these kinds of multimer moves usually have low acceptance rates, because moving a group of $(CN_bC)$ units together also produce a higher likelihood of disrupting interactions that are favorable, while introducing a large number of unfavorable collisions at the same time. The number of favorable interactions interrupted and unfavorable interactions introduced when moving a randomly selected block of $(CN_bC)$ multimer generally grows with the size of the block and limits the advantage of such random large-scale moves. To reap the benefits of large-scale MC moves without suffering impossibly high rejection rates, a stochastic potential switching (SPS) method[55] was employed. SPS is effective in accelerating MC sampling by grouping subunits of the system into naturally formed blocks that have strong interactions within each block but weak interactions among them. Using the SPS algorithm to identify and then move these blocks generate higher acceptance rates compared to randomly chosen blocks. Using the SPS algorithm formulated in [55], base stacking and pairing interaction terms were used to divide bases that have strong interactions among them into stochastically generated blocks. Under this implementation of SPS, nucleobases that are stacked or paired generally have a higher probability of being grouped together. A SPS move is then completed by applying a random translation+rotation to each block, one at a time, but using the rest of the free energy terms to accept or reject according to Metropolis. SPS strictly satisfies detailed balance[55], and the SPS blocking decisions are stochastic so two SPS moves based on the same set of stacking and pairing interactions can generate different blocking outcomes. Finally, the SPS blocks are determined stochastically based on free energy terms in the current configuration *before* the move, and since these base-base interactions evolve during the simulation, the sets of $(CN_bC)$ units that SPS decides to block together also change from one MCS to the next.

## C. Closure solutions computed on-the-fly versus precomputed solutions stored on disk

The closure solutions $C(\psi_j, \psi_{j+1}; v_j^*, v_{j+1}^* | \text{RETO}_j)$ can be computed on-the-fly during the simulation according to the prescription in Sect. II A and B. The RETO variables, which are shown in Fig, 1(e), are functions of the relative position and orientation between two sequence neighbor bases. Each time a base $j$ is moved, the RETO variables on either side of it are also moved, and the closure solutions on both the $5'$ and $3'$ sides of $j$ are recomputed. This calculation is the most computationally demanding part of the simulation.

To improve computational efficiency, the closure solutions $C(\psi_j, \psi_{j+1}; v_j^*, v_{j+1}^* | \text{RETO}_j)$ were precomputed and stored as a function of $\text{RETO}_j$. This was done by representing $C(\psi_j, \psi_{j+1}; v_j^*, v_{j+1}^* | \text{RETO}_j)$ as a matrix on a discretized grid of $(\psi_j, \psi_{j+1})$ values. The elements of this matrix are either 1 or 0, so they could be stored in bit-encoded sparse matrix format and retrieved using low-level bit arithmetic functions. These $C$ matrices were then saved for every possible pair of $(v_j^*, v_{j+1}^*)$ values, also on a discretized grid. The entire set of $C$ matrices were then computed for each set of $\text{RETO}_j$ values, again on a discretized grid, typically 64×64×64×64. For RETO values inside this grid, the simulation looked up the $C$ matrices from the precomputed library, whose size was typically ~10 gigabytes. For RETO values outside the grid, the simulation computed $C$ on the fly. Using these $C$ matrices, the integrals in Eq.(1) were approximated by matrix multiplications. During the simulation, the library lookup and data retrieval was handled by a separate server program running independently from the MC. The MC made requests to the server through a named pipe or a first-in/first-out (FIFO) queue, and the server returned the requested data to the MC. In this way, one server can serve multiple instances of several independent MC simulations running simultaneously. For on-the-fly calculations of $C$, openMP was used to parallelize some of the loops to further speed up the program, using 1, 2 or 4 cores depending on machine architecture. Leveraging all these numerical strategies, a DNA could be folded from scratch starting from an open conformation in about 100,000 MCS, which typically took about 1 to 2 weeks running single-threaded on an AMD EPYC 7281 processor.

## IV. Results and Discussion

CCMC simulations were used to fold two inverted repeat DNA sequences, $(dA)_{11}$-$(dG)_4$-$(dT)_{11}$ and $(dG)_{11}$-$(dA)_4$-$(dC)_{11}$. The optimally folded structure for each should result in an 11-bp helix, which is about one complete turn, plus a 4-nt hairpin loop. On the genome, inverted repeat (IR) sequences like these are known to be able to form cruciform structures[12–14,16,56]. But since these two particular IRs have many possible ways to make base pairs among different nucleotides on the sequence, a perfect 4-bp hairpin with an 11-bp helix is unlikely to be the only relevant conformation at equilibrium. Some of the simulated structures show frayed end(s) with dangling unpaired bases on the helix, while others folded into triplexes and quadruplexes. But the majority of them folded into well-defined B-form helices.

Fig. 5 shows results of 30 independent folding simulations of the GC IR, $(dG)_{11}$-$(dA)_4$-$(dC)_{11}$, each having started from a different open conformation. The graph shows the value of the free energy cost function $\Delta F$ for each structure versus its RMSD relative to a perfect B-form helix. While there are a hairpin DNA structures of various lengths in the crystallographic database, we have measure RMSD using the 11-nt fragments on both the 5′ and 3′ ends of the GC IR, against the two strands in the double-stranded DNA helix in the Drew-Dickerson dodecamer [57] (PDB ID: 1BNA). Measuring RMSD this way, the measurement is both sensitive to the presence of any dangling ends on the hairpin, as well as any bending along the axis of the helix. All of the folded structures had right-handed twists. Their final folded structures are shown in Fig. 5, labeled 1 through 9 and a through u, corresponding to the points on the graph. All the structures shown in Fig. 5 with RMSD less than 4 Å folded into a single helix, with varying lengths of frayed ends or minor imperfections. These account for about half of the 30 folding simulations. Some of the structures with the best RMSD are 2, j, 4, a, 8, 1, k, i, m, 3, b and l. These structures also have the lowest cost function values in the free energy model. The number of strong ($< $ -8 kcal/mol, red circles), medium (between –8 and -4 kcal/mol, orange squares) and weak (between –4 and –2 kcal/mol, blue triangles) base pair interactions and which pair of bases were involved in each are shown in the form of an adjacency matrix[58,59] next to each structure. In each matrix, the sequence index increases from bottom to top on the vertical axis, and from left to right on the top axis, and base pairs are only shown on the upper half of the matrix. In this adjacency map, a perfectly formed 11-bp helix in the GC IR structure should have base pairs reflected by a red diagonal line extending from the top left to the bottom right on the upper half plane of the matrix. Structures with the best RMSD all show base pairs along or parallel to this diagonal line on the adjacency map, indicating that they all folded into helices . The rest of the structures with RMSD greater than approximately 4 Å did not fold into helices. Instead, many of these exhibited strong triplex DNA interactions (e.g. structures 7, 9, e, n, r, s and t). Based on the value of their cost functions $\Delta F$, these appear to be metastable structures caught in misfolded conformations, which the simulations were unable to fully anneal into more stable folds. Overall, the folding results displayed in Fig. 5 are encouraging. The cost function $\Delta F$, which is based on the theoretical free energy model described in Sect. II, appears to be able to distinguish correctly folded structures from misfolded ones, and the MC algorithm described in Sect. III seems to have sufficient numerical efficiency to fold an open GC IR sequence *ab initio* into a hairpin with the expected structure helix in 15 out of 30 simulations, within practically accessible CPU time.

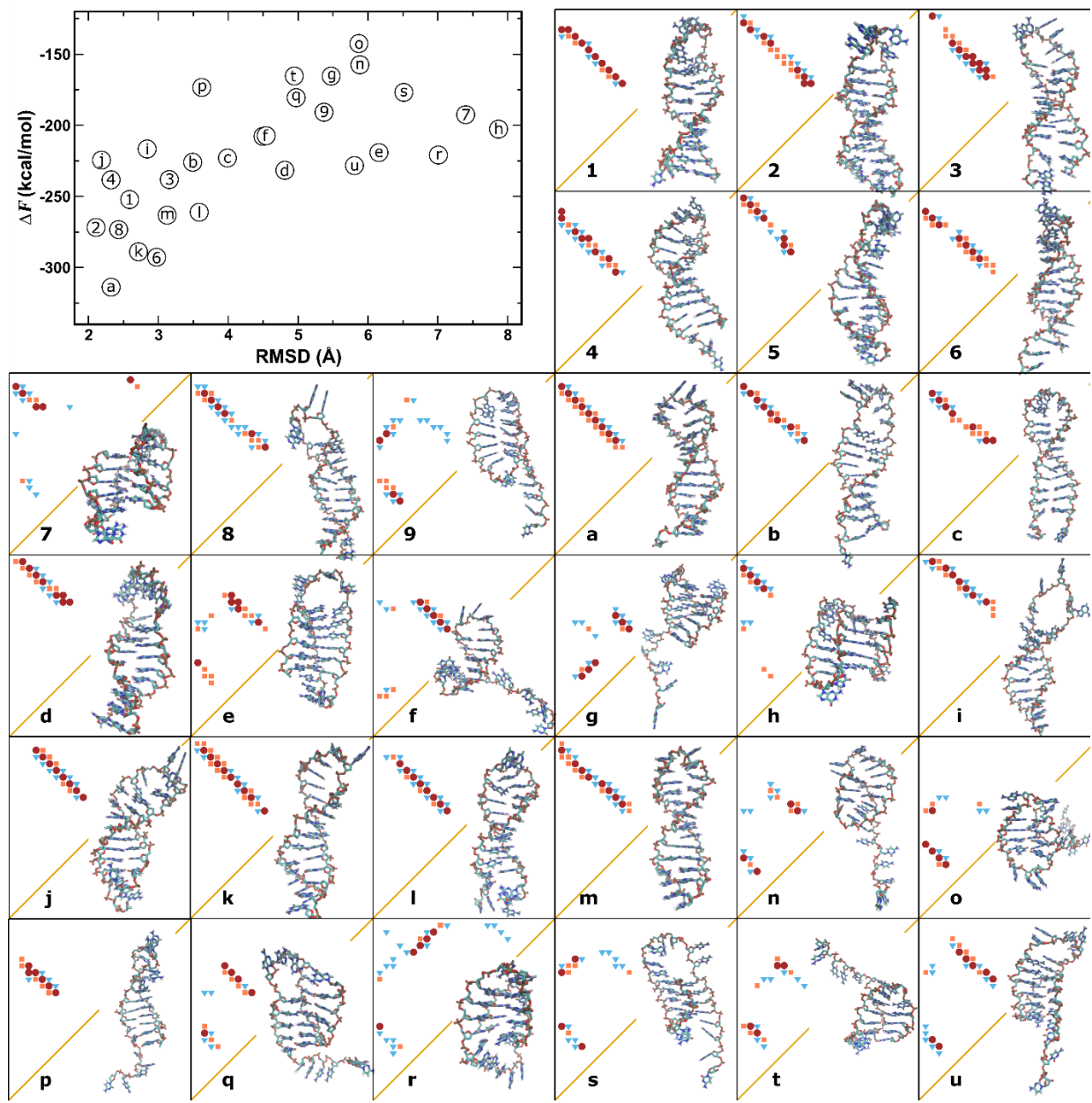

**FIGURE 5.**

Results from thirty independent folding simulations for a GC inverted repeat, $(dG)_{11}$-$(dA)_4$-$(dC)_{11,}$ after 100000 MCS. Graph shows RMSD of the 11-nt segments on the 5′ and 3′ ends for each final structure relative to a double-stranded B-DNA, and the value of the cost function $\Delta F$ from the free energy model described in Sect. II. The base pairs in each structure and their strengths (red circles = < -8 kcal/mol, orange squares = between -8 and -4 kcal/mol, blue triangles = between -4 and -2 kcal/mol) are shown for each structure in the form of an adjacency matrix map.

Figs. 6 and 7 show two sample MC trajectories from these folding simulations. The trajectory in Fig. 6 corresponds to structure 1 in Fig. 5, and that in Fig. 7 to structure 2. In both Figs. 6 and 7, a MC time stamp is shown on each frame, and like Fig. 5, a corresponding base pair map is also shown on the upper left half plane. Both simulations were carried out to 200,000 MCS.

In the early part of the MC trajectory in Fig. 6, the chain seems to transition among a vast number of conformations exploring different alternative base pairing schemes, but without committing to a particular one. The base pair maps in Fig. 6 suggest that the pairing scheme was fluid, until about 6000 MCS, at which point a few persistent base pairs began to emerge. Notice that these base pairs that were formed early, e.g. those between MCS 7000 and 50000, led to a fairly long dangling dC fragment on the 3′ end of the chain. But the base pair maps show that the base pairs in the helix progressively shifted to form more base pairs, while simultaneously reducing the length of the dangling dC sequence. At around 130000 MCS, the dangling fragment was reduced to 2 nt, and at around 180000 MCS, it was further reduced to 1 nt. However, it is clear from the simulations that the evolution in the base pairing scheme was stochastic – there was no overall directionality. For example, the configurations at 190000 and 200000 MCS were less optimal than those at 170000 and 180000 MCS. The trajectory in Fig. 6 suggests that the folding appears to proceed from the initially open chain into what looks like a "molten globule" phase, where various base pairing schemes were explored, proceeding onto one metastable fold, which finally, through extended annealing, moved through a progression of more optimally folded states, toward the most optimally folded conformation, which correspond to a 11-bp helix with a 4-nt hairpin loop.

The MC trajectory in Fig. 7 tells a somewhat similar story to Fig. 6. In Fig. 7, a metastable structure was apparently seeded at approximately 20000 MCS, which led to the metastable structure at 100000 MCS, with a 3-nt dangling dC segment on the 3′ end and a 2×0 asymmetric interior bulge on the middle of the helix. Once the folding has committed to this base pairing scheme, both the dangling end and the asymmetric bulge were gradually annealed away beyond MCS 100000, producing the near-optimal hairpin-helix structure observed at 200000 MCS. Although there was no overall directionality in the folding due to its stochastic nature, the results in Figs. 5 through 7 suggest that there appear to be a folding funnel in the case of this GC IR, steering the folding process toward the maximally optimal fold. For those runs shown in Fig. 5 that did not produce helices at the end of 100000 MCS, every one of them was trapped in a metastable fold. Presumably, given a sufficiently long MC run, the simulation may be able to anneal these trapped structures to arrive at a more optimal fold, but the ergodicity of the MC moves used in the present simulation appears to be inadequate to free those misfolded conformations from their metastable states, even after being run to 200000 MCS. Annealing these structures by cycling the temperature may possibly move them toward the optimally folded state. This is currently being explored.

For the AT IR, $(dA)_{11}$-$(dG)_4$-$(dT)_{11}$, the same set of simulations were carried out, but none of the 30 runs folded into the expected helix structure. The majority of the runs folded into random globular structures with what appeared to A|T pairs, but their pairing was predominantly non-Watson-Crick. The MC algorithm itself should not be the issue since the same algorithm was able to fold the GC IR. The stability of an AT helix should not be the problem either, because even though A|T pairs have weaker stability than G|C, AT-rich IRs are known to be able to fold into a hairpin[60]. The culprit is most likely in the free energy model. While the model correctly represented the key physical forces for structures with G|C pairs, the model apparently failed for those with A|T pairs. We have tried to optimize all the adjustable parameters in the model, including the counterion screening length $\Lambda_{CI}$ and the counterion dielectric $\epsilon_{CI}$ in Eq.(2), as well as the attenuate parameter $\epsilon_{AT}$ for the *ad hoc* base pairing model to produce stable folded helices for the AT IR, but so far we have not found a set of parameters that could correctly fold it. Since the only difference between the GC IR, which the CCMC simulation was able to fold, and the AT IR is in the *ad hoc* base pairing model, we tried to search for an optimal value of $\epsilon_{AT}$ by performing a large number of parallel folding simulations for the AT IR using $\epsilon_{AT}$ ranging from 4.2 (the optimal value found for $\epsilon_{GC}$) to 1.5. None of these folded into the AT IR into a helix. These observations suggest that while the *ad hoc* base pairing model described in Sect. IIE was sufficient to describe G|C pairs, it was insufficient for A|T pairs. Given the crudeness of the *ad hoc* pairing model, there are good reasons for why it does not work for A|T pairs. First, in the explicit-solvent umbrella free energy calculation[8] that generated the data in Fig. 4, it was found that the near-field water molecules, those that are within 3 Å from any Watson-Crick donor or acceptor, play a dominant role in renormalizing the direct hydrogen bonding interactions between A and T. The effects of these near-field waters are missing in the *ad hoc* pairing model. Second, there are two direct hydrogen bonds in a A|T Watson-Crick pair, but the $O_2$ atom on T interacts strongly with the $C_2$ atom on A on the minor groove side, which has a large positive partial charge ($+0.5716e$) in Amber. The $C_2$ on A and the $O_2$ atom on T tend to interact too strongly with each other in the absence of any water between them. Finally, both experiments[57] and simulations[61,62] suggest that on a DNA helix, water molecules are strongly bound to the minor groove between A|T pairs. These near-field waters on the minor groove, whose effects are missing in our *ah doc* pairing model, seem to play a particular important role in defining the A|T pairing interaction. Work is currently ongoing to develop a physically accurate pairing model for A|T, as well as G|C pairs.

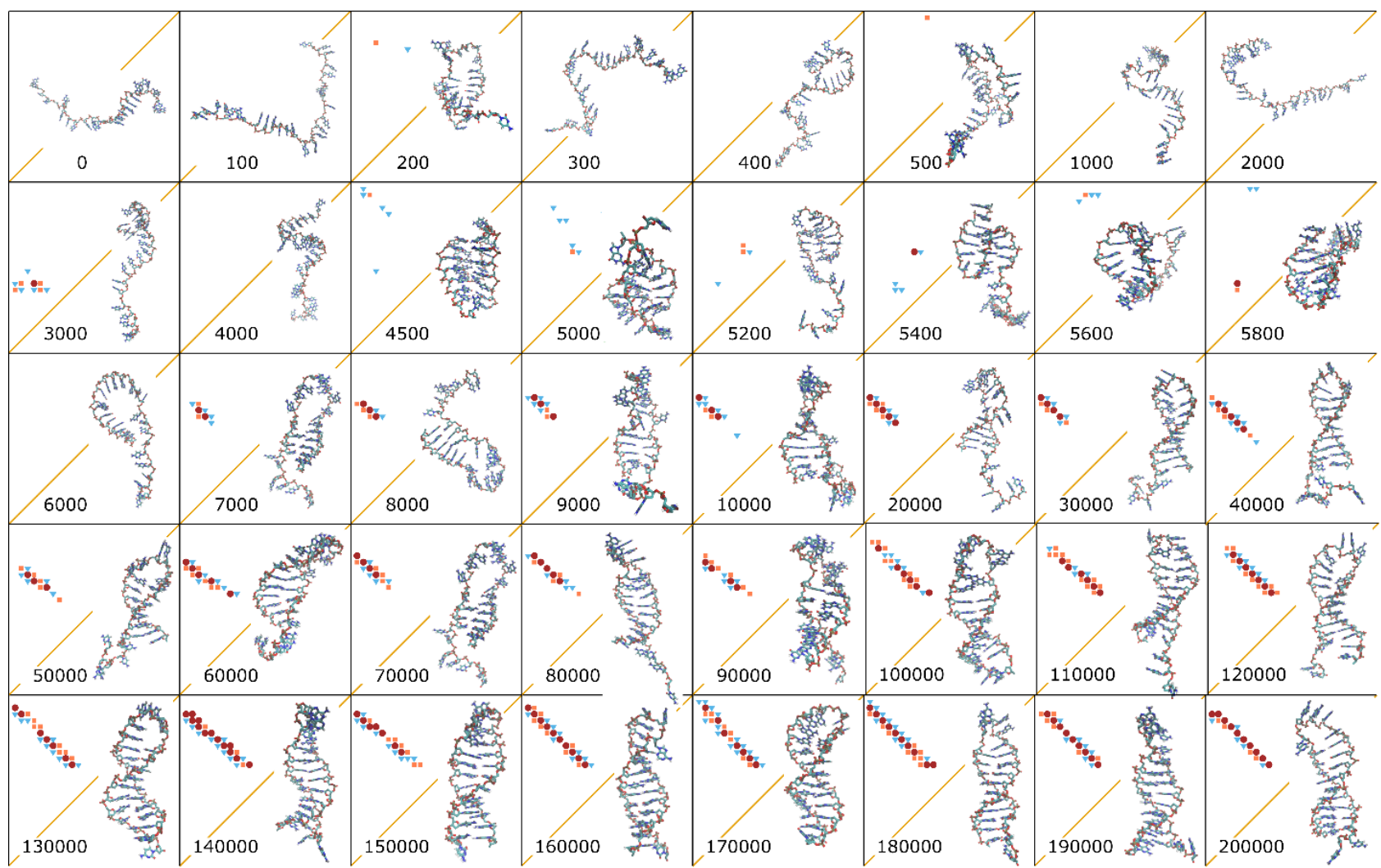

**FIGURE 6.**

A sample MC trajectory from the GC IR folding simulations, corresponding to structure 1 in Fig. 5.

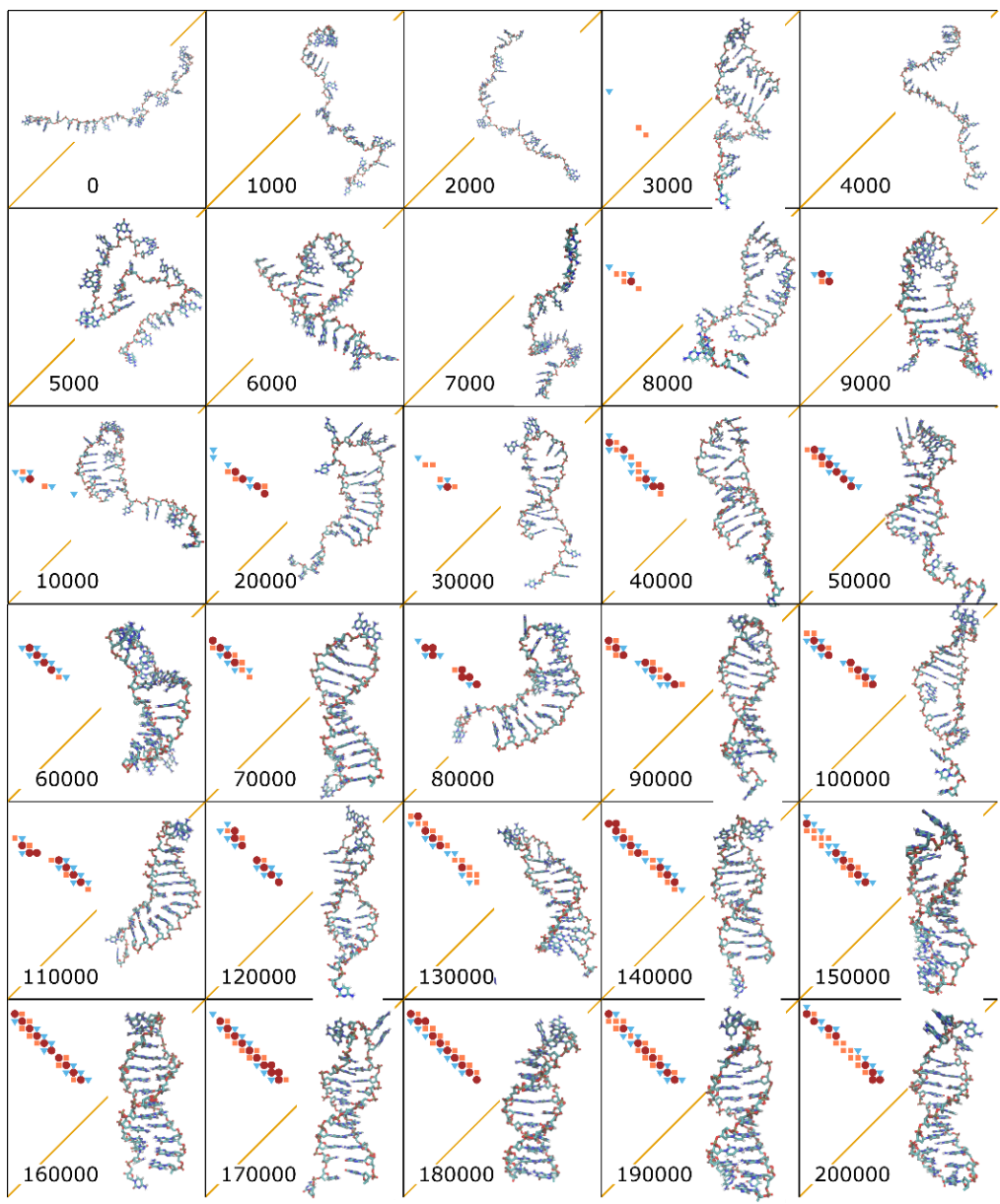

**FIGURE 7.**

Another sample MC trajectory from the GC IR folding simulations, corresponding to structure 2 in Fig. 5.

# V. Conclusion

A simulation that is able to capture the full atomistic details of nucleic acids has been used to fold two oligonucleotide inverted repeat DNA sequences from scratch. The theoretical model behind this simulation accounts for four physical molecular driving forces that are essential for folding nucleic acids. The simulation used chain closure to describe (1) the conformational entropy of the sugar-phosphate backbone, and it employed analytical theories to describe the statistical mechanics of the molecular forces behind: (2) base stacking, (3) base complementarity and (4) the counterion-induced attractive self-interactions along the sugar-phosphate backbone. While some of these molecular driving forces are predicated on water and ions in the solvent, these theories enabled the folding simulations to be carried out without adding explicit water or ions. Numerical optimizations and special MC moves were also used to accelerate the simulations, which were able to fold a GC-rich inverted repeat DNA sequence into the expected hairpin structure with a canonical B-form helix in moderate CPU time. The results of the simulations demonstrated that the four key molecular driving forces included in the theoretical model were both adequate and essential for folding nucleic acids. The results also suggest that the folding of DNAs appear to proceed via what looks like a "molten globule" phase, where different alternative base pairing schemes are actively explored by the DNA, proceeding onto one committed metastable fold, which finally, through extended annealing, moves through a progression of more stable folds, toward the most optimally folded conformation. While there was no absolute directionality in the folding trajectories, there appears to be a folding funnel in the case of the GC-rich inverted repeat, steering the folding process toward the maximally optimal fold. Some of the other folding trajectories that did not produce the expected hairpin structure at the end of the simulation appear to be caught in a metastable state, many exhibiting strong triplex interactions, while in other trajectories, initially misfolded structures were able to free themselves out of metastable states to move closer to a globally optimal fold, after extended equilibration.

When applied to an AT-rich inverted repeat DNA of the same length, the simulations were unable to fold it into any regular helix or hairpin structure. Experimentally, AT-rich inverted repeats have been shown to form hairpins in solution, and since every aspect of the simulations was identical between the two sequences except in the theoretical model for the pairing interactions, the benchmark studies in this paper suggest that the simplistic *ad hoc* pairing model employed in the simulation was unable to capture the physics of A|T complementarity to correctly fold DNAs with A|T base pairs, and a physically accurate pairing model is needed. However, these simulations also demonstrated that when the four key physical molecular driving forces (backbone entropy, base stacking, base pairing and counterion-induced self-interactions) are in place, nucleic acids can be folded *ab initio*, provided that a computationally effective Monte Carlo sampling algorithm is implemented at the same time.

## Acknowledgement

This material is based upon work supported by the National Science Foundation under Grant No. CHE-1664801.